\documentclass[aps,prd,twocolumn,epsf,nofootinbib,superscriptaddress]{revtex4}

\usepackage{graphicx,here}
\usepackage{amsmath,amssymb,latexsym}
\usepackage{bm}

\def\lsim{\mathrel{\raise.3ex\hbox{$<$\kern-.75em\lower1ex\hbox{$\sim$}}}}
\def\gsim{\mathrel{\raise.3ex\hbox{$>$\kern-.75em\lower1ex\hbox{$\sim$}}}}

\def\lbldef#1#2{\expandafter\gdef\csname #1\endcsname {#2}}

\def\href#1#2{#2}

\newcommand{\bwide}{\begin{widetext}}
\newcommand{\ewide}{\end{widetext}}
\newcommand{\beq}[1]{\begin{equation} \label{(#1)}}
\newcommand{\eeq}{\end{equation}}
\newcommand{\ba}[1]{\begin{eqnarray} \label{(#1)}}
\newcommand{\ea}{\end{eqnarray}}

\begin{document}
\hspace*{130mm}{\large \tt FERMILAB-PUB-09-494-A}

\title{Possible Evidence For Dark Matter Annihilation In The Inner Milky Way From The Fermi Gamma Ray Space Telescope}

\author{Lisa Goodenough}
\affiliation{Center for Cosmology and Particle Physics, Department of Physics, New York University, New York, NY  10003}
\author{Dan Hooper}
\affiliation{Center for Particle Astrophysics, Fermi National Accelerator Laboratory, Batavia, IL 60510}
\affiliation{Department of Astronomy and Astrophysics, University of Chicago, Chicago, IL  60637}

\begin{abstract}

We study the gamma rays observed by the Fermi Gamma Ray Space Telescope from the direction of the Galactic Center and find that their angular distribution and energy spectrum are well described by a dark matter annihilation scenario. In particular, we find a good fit to the data for dark matter particles with a 25-30 GeV mass, an annihilation cross section of $\sim 9\times 10^{-26}$ cm$^3$/s, and that are distributed with a cusped halo profile, $\rho(r) \propto r^{-1.1}$, within the inner kiloparsec of the Galaxy. We cannot, however, exclude the possibility that these photons originate from an astrophysical source or sources with a similar morphology and spectral shape to those predicted in an annihilating dark matter scenario.

\end{abstract}


\maketitle


Searches for dark matter annihilation products are among the most exciting missions of the Fermi Gamma Ray Space Telescope (FGST). In particular, the FGST collaboration hopes to observe and identify gamma rays from dark matter annihilations occuring cosmologically~\cite{cosmo}, as well as within the Galactic Halo~\cite{halo}, dwarf galaxies~\cite{dwarf}, microhalos~\cite{micro}, and the inner region of the Milky Way~\cite{inner}. 

Due to the very high densities of dark matter predicted to be present in the central region of our galaxy, the inner Milky Way is expected to be the single brightest source of dark matter annihilation radiation in the sky. This region is astrophysically rich and complex, however, making the task of separating dark matter annihilation products from backgrounds potentially challenging. In particular, the Galactic Center contains a $2.6\times 10^6\, M_{\odot}$ black hole coincident with the radio source Sgr A$^*$~\cite{ghez}, the supernova remnant Sgr A East, and a wide variety of other notable astrophysical objects, including massive O and B type stars, and massive compact star clusters (Arches and Quintuplet). 

Since its launch in June of 2008, the Large Area Telescope (LAT) onboard the FGST has identified as photons several hundred thousand events from within a few degrees around the Galactic Center. In addition to possessing the effective area required to accumulate this very large number of events, the angular resolution and energy resolution of the FGST's LAT are considerably improved relative to those of its predecessor EGRET. As a result, this new data provides an opportunity to perform a powerful search for evidence of dark matter annihilation~\cite{fermidark}.

Dark matter annihilations are predicted to produce a distribution of gamma rays described by:
\begin{equation}
\Phi_{\gamma}(E_{\gamma},\psi) =\frac{1}{2} <\sigma v> \frac{dN_{\gamma}}{dE_{\gamma}} \frac{1}{4\pi m^2_{\rm{dm}}} \int_{\rm{los}} \rho^2(r) dl(\psi) d\psi,
\label{flux1}
\end{equation}
where $<\sigma v>$ is the dark matter particle's self-annihilation cross section (multiplied by velocity), $m_{\rm dm}$ is the dark matter particle's mass, $\psi$ is the angle away from the direction of the Galactic Center that is observed, $\rho(r)$ describes the dark matter density profile, and the integral is performed over the line-of-sight. $dN_{\gamma}/dE_{\gamma}$ is the spectrum of prompt gamma rays generated per annihilation, which depends on the dominant annihilation channel(s). Note that Eq.~\ref{flux1} provides us with predictions for both the distribution of photons as a function of energy, and as a function of the angle observed. It is this powerful combination of signatures that we will use to identify and separate dark matter annihilation products from astrophysical backgrounds~\cite{method}.

With a perfect gamma ray detector, the distribution of dark matter events observed would follow precisely that described in Eq.~\ref{flux1}. The LAT of the FGST has a finite point spread function, however, which will distort the observed angular distribution. In our analysis, we have modeled the point spread function of the FGST's LAT according to the performance described in Ref.~\cite{performance}.


In addition to any gamma rays from dark matter annihilations coming from the region of the Galactic Center, significant astrophysical backgrounds are known to exist. In particular, HESS~\cite{hess} and other ground-based gamma ray telescopes~\cite{other} have detected a rather bright gamma ray source coincident with the dynamical center of our galaxy ($l=-0.055^{\circ}$, $b=0.0442^{\circ}$). The spectrum of this source has been measured to be a power-law of the form $dN_{\gamma}/dE_{\gamma} \approx 10^{-8} \, {\rm GeV}^{-1}\, {\rm cm}^{-2}\, {\rm s}^{-1} (E/{\rm GeV})^{-2.25}$ between approximately 160 GeV and 20 TeV. Although HESS and other ground-based telescopes cannot easily measure the spectrum of this source at lower energies, it is likely that it will extend well into the range studied by the FGST~\cite{dermer}, where it will provide a significant background for dark matter searches~\cite{gabi}. Furthermore, gamma ray emission from numerous faint point sources and/or truly diffuse sources can provide a formidable background in the region of the Galactic Center, especially near the disk of the Milky Way. To model this background, we have studied the angular distribution of photons observed by the FGST in the region of $3^{\circ}< |l| <6^{\circ}$, and found that the emission is fairly well described by a function which falls off exponentially away from the disk with a scale of roughly $2.2^{\circ}$ to $1.2^{\circ}$ for photons between 300 MeV and 30 GeV, respectively. In addition to the previously described HESS point source, we will use this angular distribution as a template to model the diffuse background in our analysis. 

To start, we consider dark matter distributed according to a Navarro-Frenk-White (NFW)~\cite{nfw} halo profile with a scale radius of 20 kpc and normalized such that the dark matter density at the location of the Solar System is equal to its value inferred by observations~\cite{ullio}. This profile, however, combined with the background model described above, leads to a distribution of photons that falls off more slowly as a function of angle from the Galaxy's dynamical center than is observed by the FGST. If we steepen the halo profile slightly, such that $\rho(r) \propto 1/r^{\gamma}$, with $\gamma=1.1$ ($\gamma=1.0$ for an NFW profile), the angular distribution of the events matches the observations of the FGST very well. Such a steepening of the inner cusp could result from, for example, the adiabatic contraction of the profile due to the dynamics of the baryonic gas which dominate the gravitational potential of the Inner Galaxy~\cite{ac}.

\begin{figure*}[!]
\begin{center}
{\includegraphics[angle=0,width=0.37\linewidth]{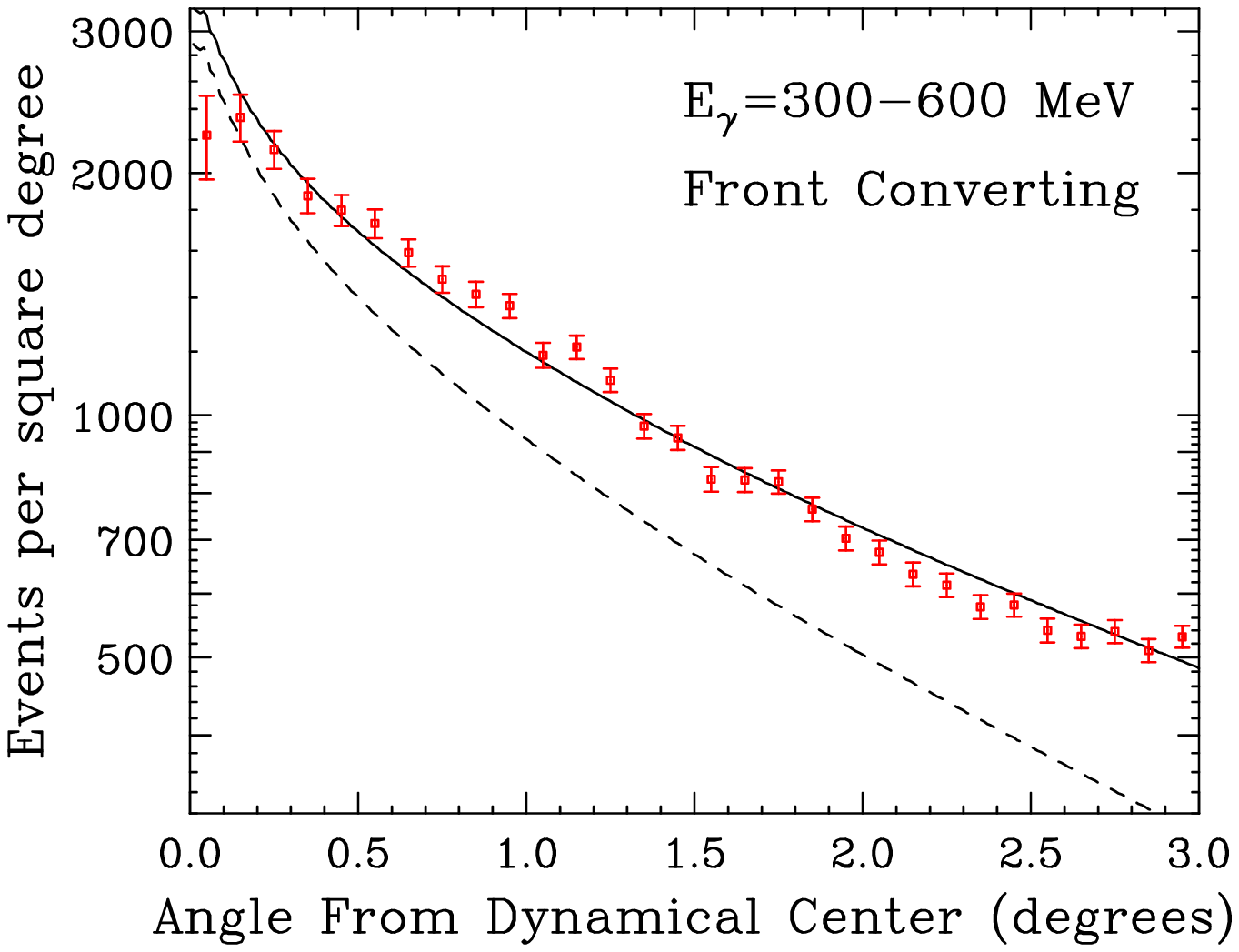}}
\hspace{0.5cm}
{\includegraphics[angle=0,width=0.37\linewidth]{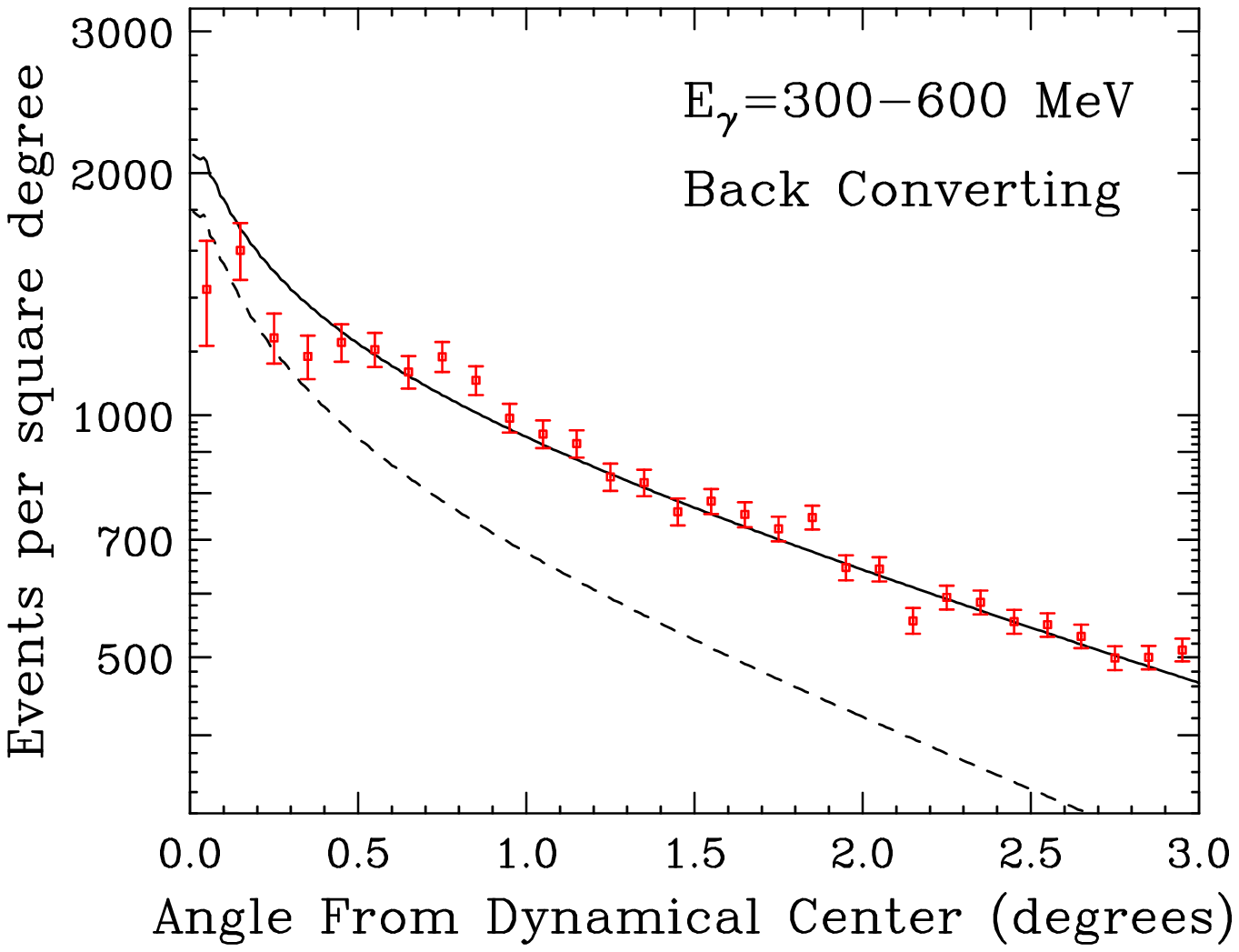}}\\
{\includegraphics[angle=0,width=0.37\linewidth]{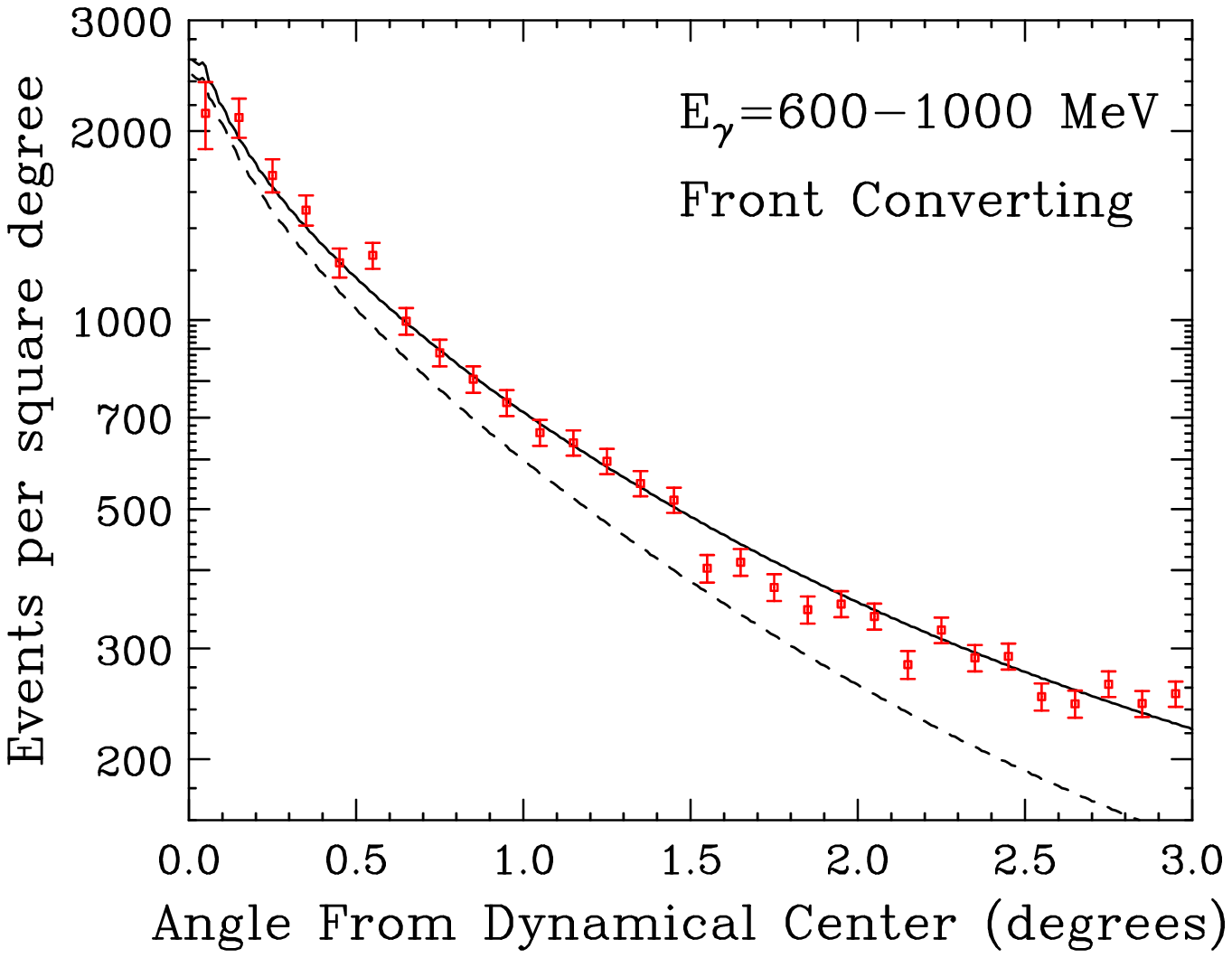}}
\hspace{0.5cm}
{\includegraphics[angle=0,width=0.37\linewidth]{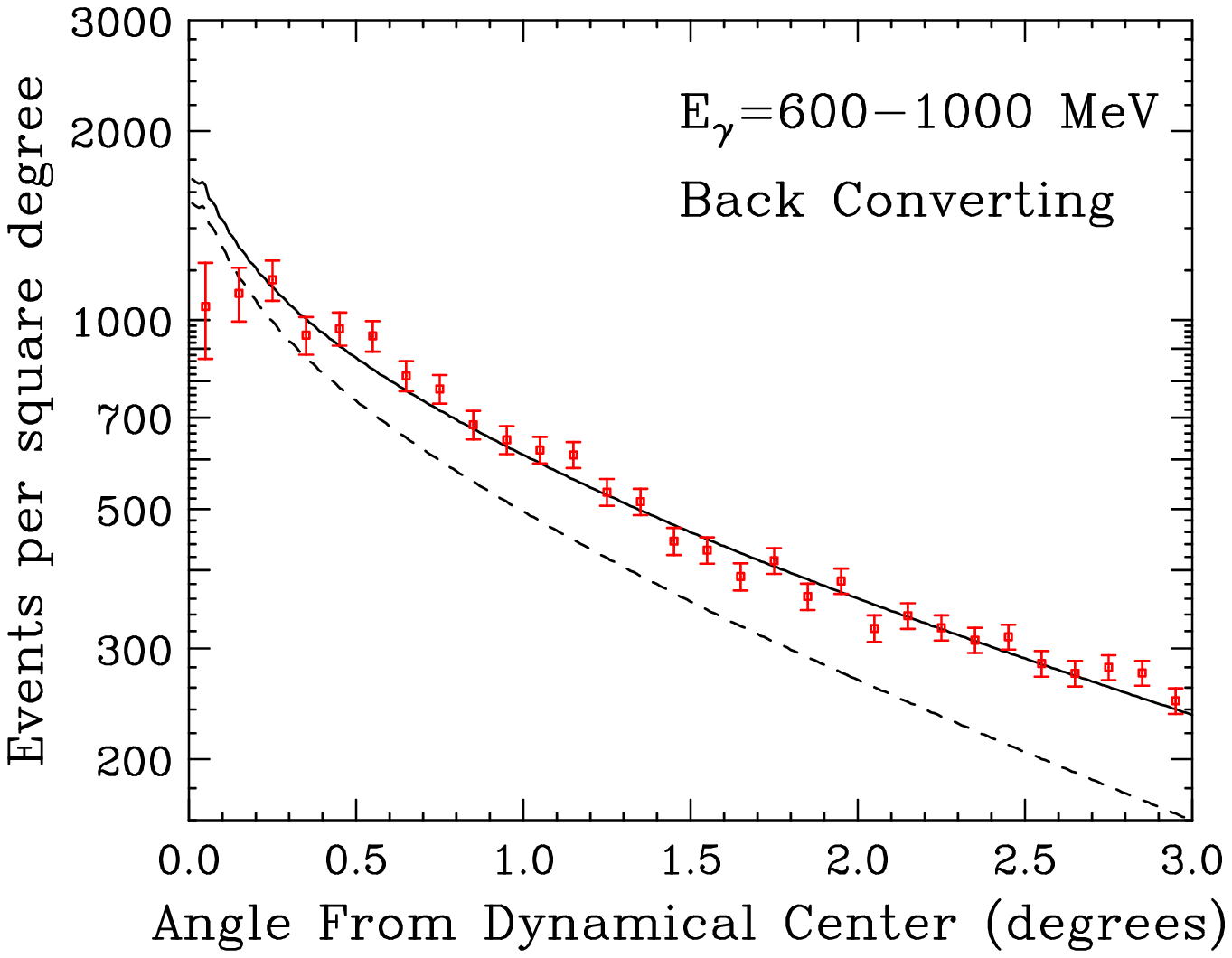}}\\
{\includegraphics[angle=0,width=0.37\linewidth]{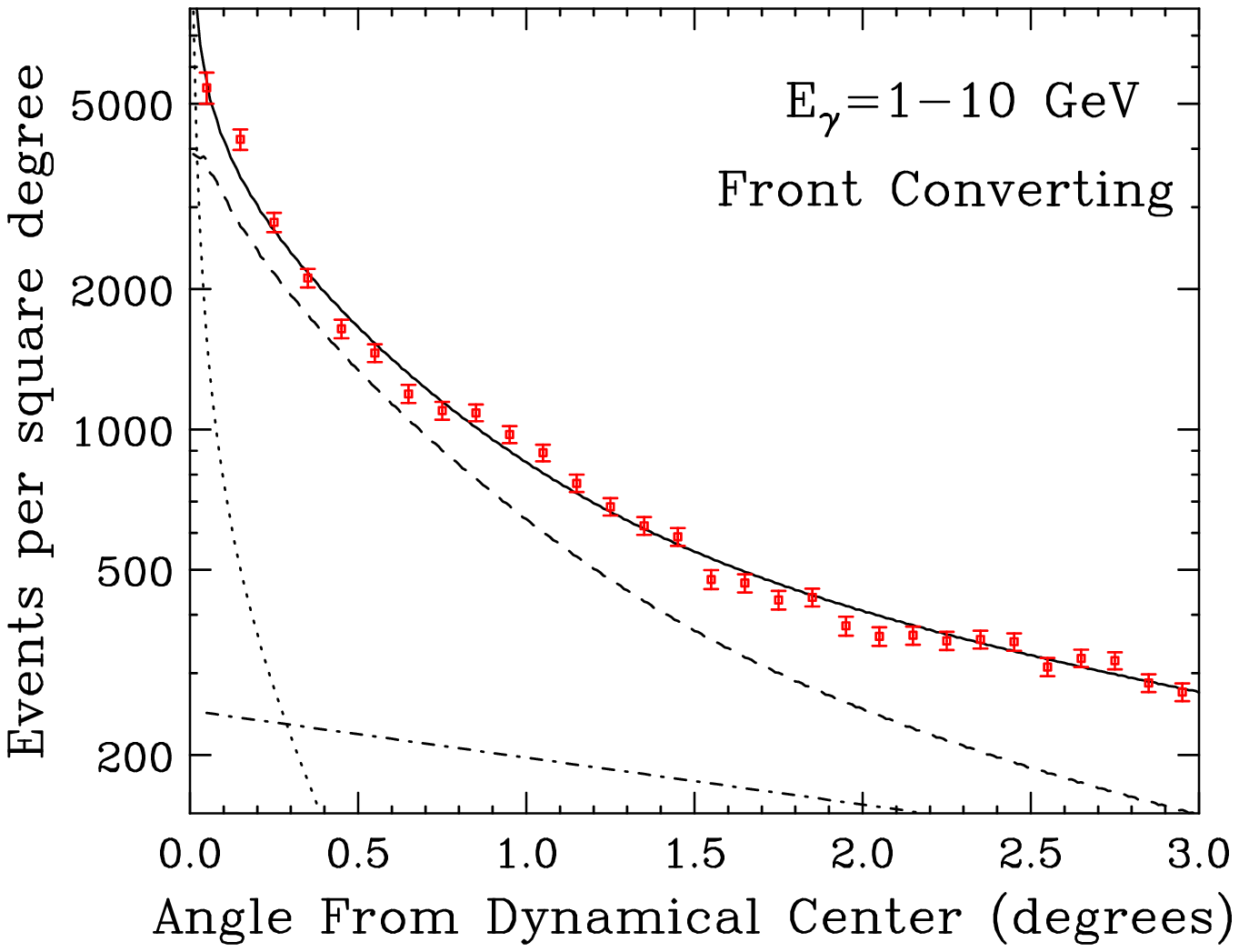}}
\hspace{0.5cm}
{\includegraphics[angle=0,width=0.37\linewidth]{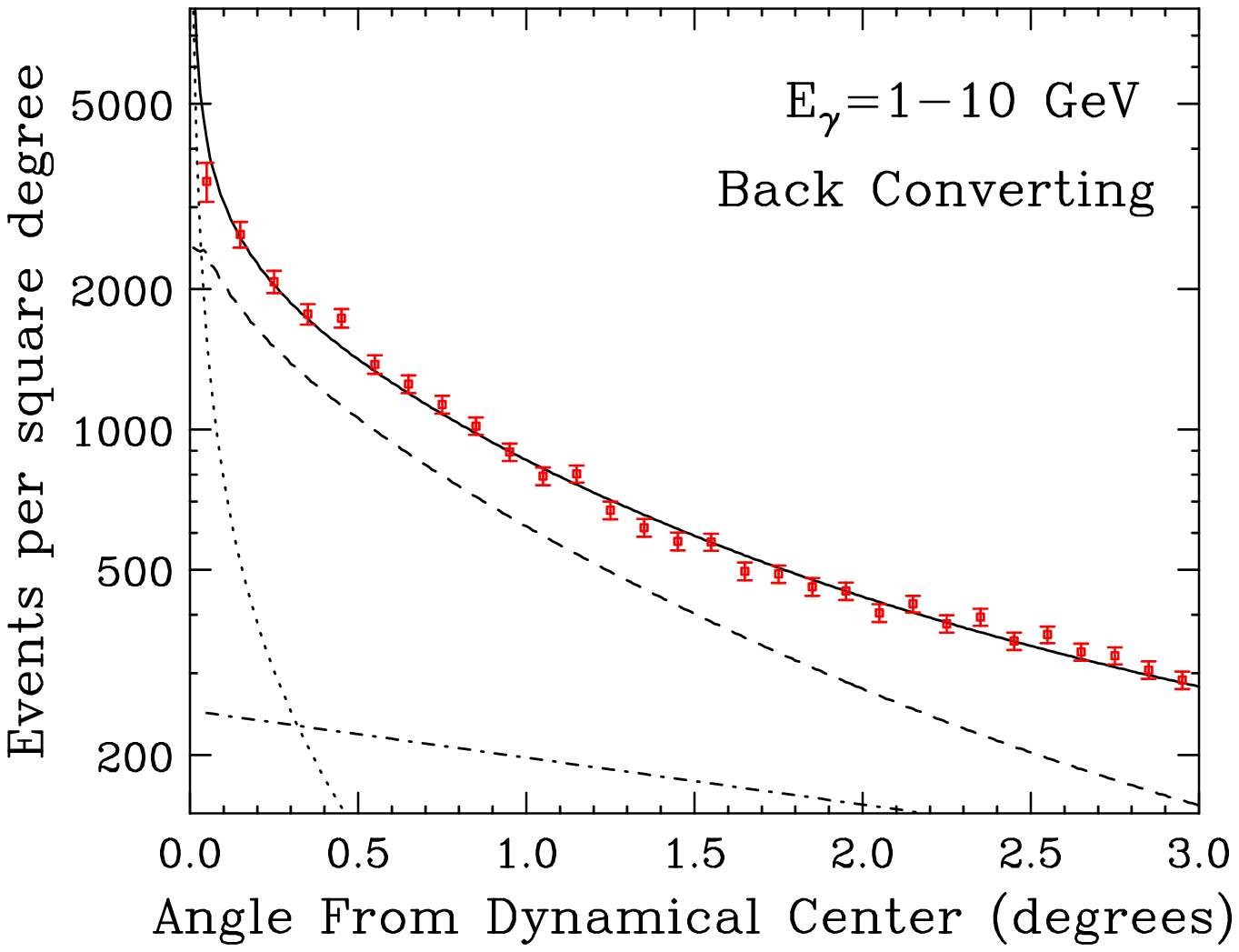}}\\
{\includegraphics[angle=0,width=0.37\linewidth]{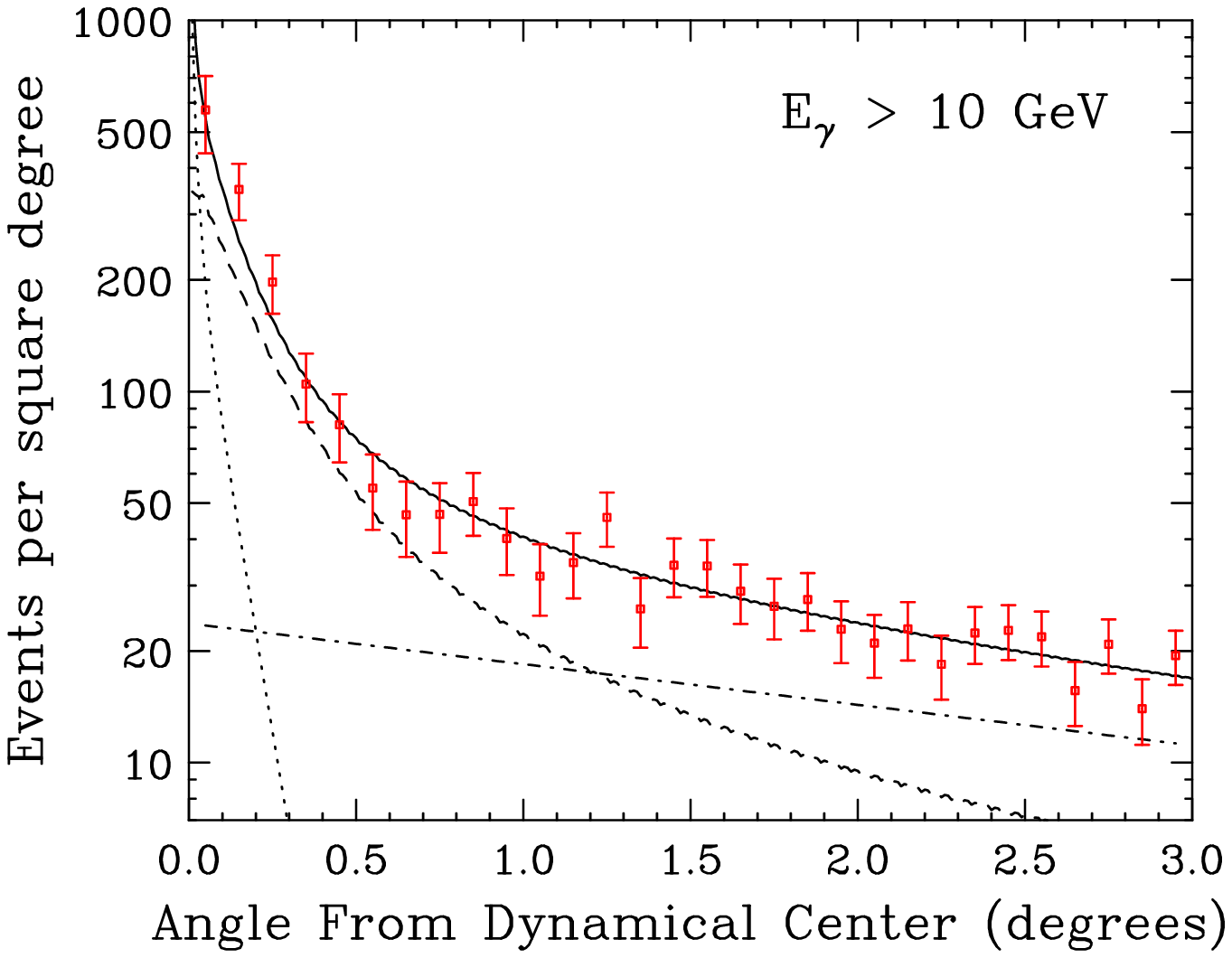}}
\hspace{0.5cm}
{\includegraphics[angle=0,width=0.37\linewidth]{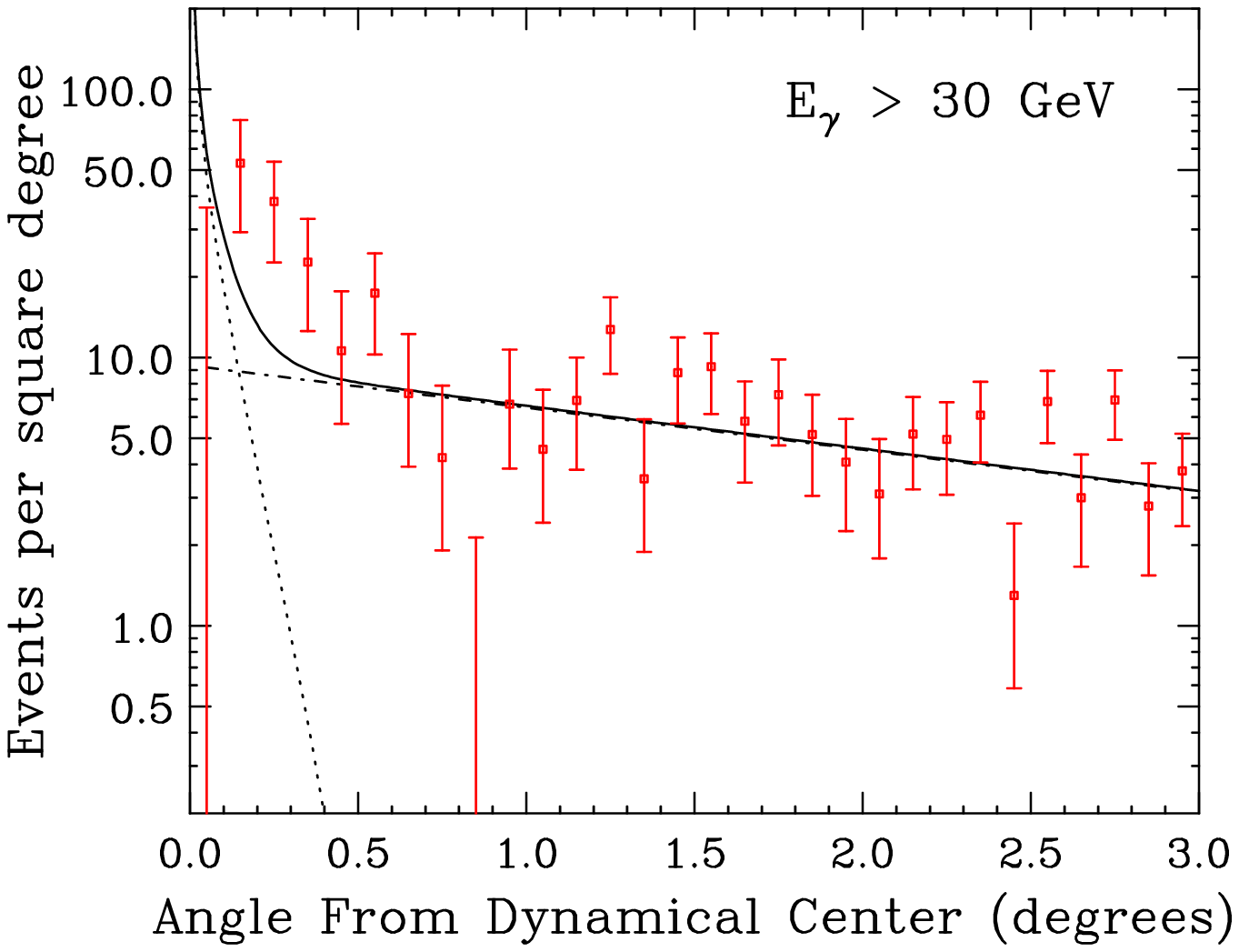}}\\
\vspace{-0.3cm}
\caption{The angular distribution of gamma rays around the Galactic Center observed by the FGST. In each frame, the dashed line denotes the shape predicted for the annihilation products of dark matter distributed according to a halo profile which is slightly cuspier than NFW ($\gamma=1.1$). The dotted line is the prediction from the previously discovered TeV point source located at the Milky Way's dynamical center, while the dot-dashed line denotes the diffuse background described in the text (which, although included in each case, falls below the range of rates shown in the upper four frames). The solid line is the sum of these contributions.}
\label{angular}
\end{center}
\end{figure*}

In Fig.~\ref{angular}, we show the angular distribution of gamma rays as a function of angle from the Galaxy's dynamical center, as measured by the FGST over the period of August 4, 2008 to October 5, 2009, corresponding to an exposure in the Galactic Center region of approximately $2.2\times 10^{10}$ cm$^2$ sec at 300 MeV, and increasing to between 3.5-4.3 $\times 10^{10}$ cm$^2$ sec between 1 and 300 GeV. We include only those events in the ``diffuse'' class (as defined by the FGST collaboration) and do not include events with zenith angle greater than 105$^{\circ}$, or from the region of the South Atlantic Anomaly. Below 10 GeV, we show separately those events that were converted in the front (thin) and back (thick) regions of the detector (these are treated separately to account for the differing point spread functions for these event classes). In each frame, we compare these measurements to the angular distribution predicted for photon from annihilating dark matter with $\gamma=1.1$, the HESS point source, the diffuse background, and for the sum of these contributions. In each case, these predictions take into account the point spread function of the FGST. For each range of energies shown, we have normalized the diffuse background and dark matter contributions to provide the best fit to the data. Above 1 GeV, the normalization of the HESS source was determined by extrapolating the measured power law spectrum to lower energies, which appears to be consistent with the distribution observed by the FGST. Below 1 GeV, the FGST data do not appear to contain significant emission from this point source, so we suppress this contribution in these energy bins under the assumption that the extrapolation of the HESS spectrum becomes invalid around $\sim 1$ GeV.

Thus far, we have performed these fits while remaining agnostic about the mass, annihilation cross section, and dominant annihilation channels of the dark matter particle, simply normalizing the angular shape predicted by our selected halo profile to the data. By comparing the relative normalizations required in each energy range shown in Fig.~\ref{angular}, however, one can infer the shape of the gamma ray spectrum from our dark matter component and for the diffuse background. In particular, the relative normalizations required of the diffuse component in the various energy ranges shown in Fig.~\ref{angular} imply that this background has a spectrum very approximately of the form $dN_{\gamma}/dE_{\gamma} \propto E^{-2.3}$.

\begin{figure*}[!]
\begin{center}
{\includegraphics[angle=0,width=0.49\linewidth]{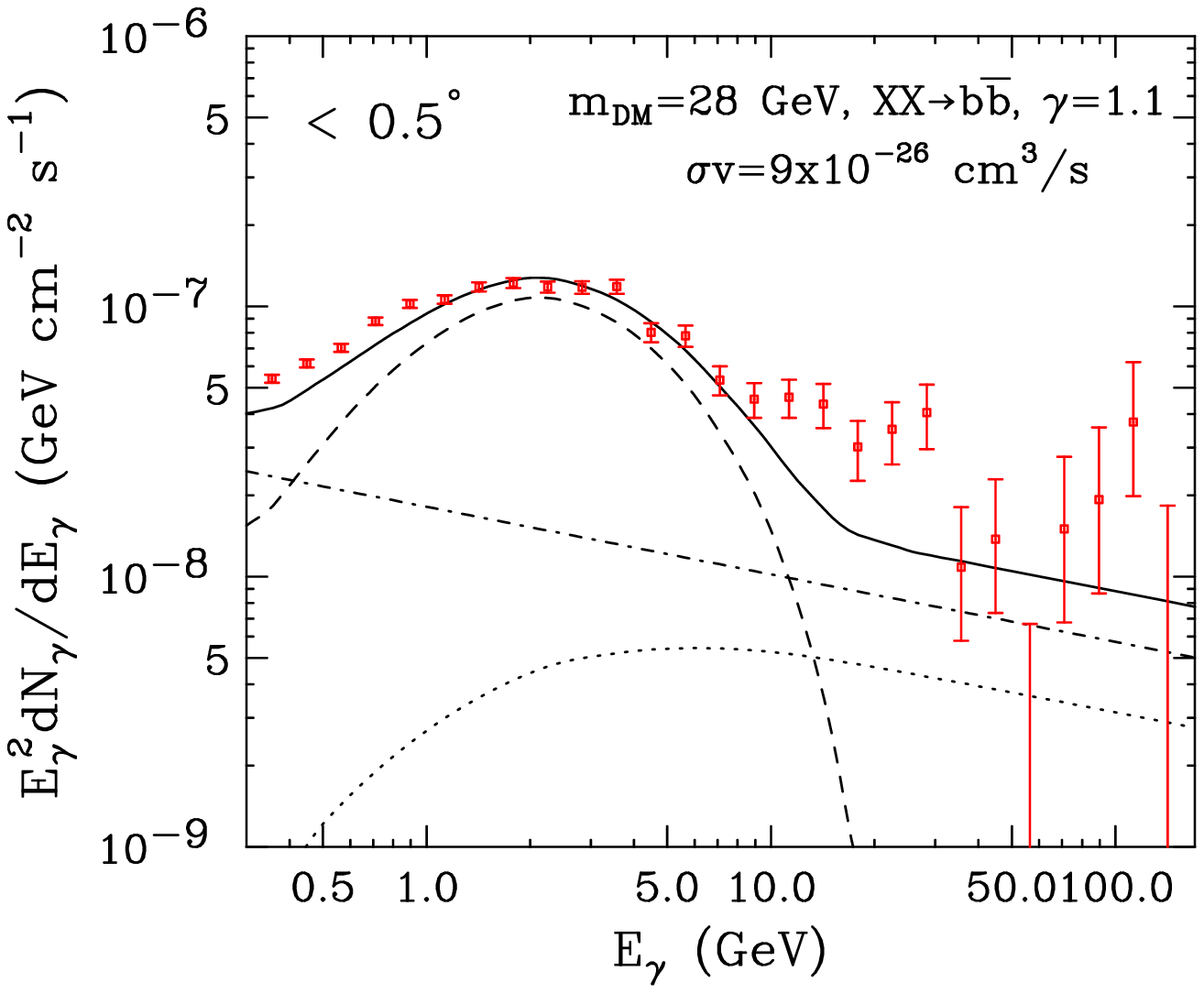}}
\hspace{0.1cm}
{\includegraphics[angle=0,width=0.49\linewidth]{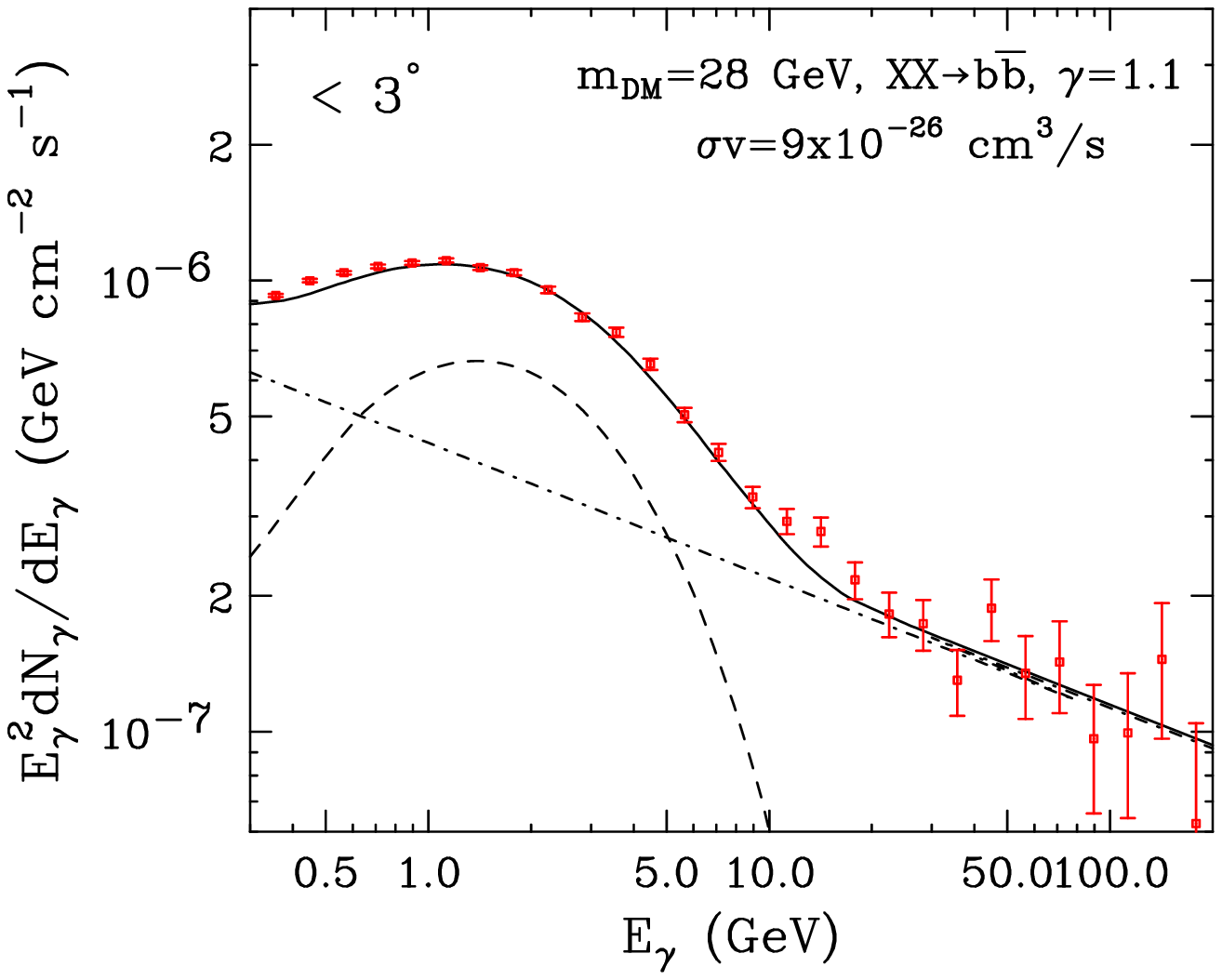}}\\
\vspace{-0.3cm}
\caption{The gamma ray spectrum measured by the FGST within 0.5$^{\circ}$ (left) and 3$^{\circ}$ (right) of the Milky Way's dynamical center. In each frame, the dashed line denotes the predicted spectrum from a 28 GeV dark matter particle annihilating to $b\bar{b}$ with a cross section of $\sigma v = 9\times 10^{-26}$ cm$^3$/s, and distributed according to a halo profile slightly more cusped than NFW ($\gamma=1.1$). The dotted and dot-dashed lines denote the contributions from the previously discovered TeV point source located at the Milky Way's dynamical center and the diffuse background, respectively.  The solid line is the sum of these contributions.}
\label{spectrum}
\end{center}
\end{figure*}

In Fig.~\ref{spectrum}, we show the spectrum of photons observed by the FGST within 0.5$^{\circ}$ and 3$^{\circ}$ of the Galaxy's dynamical center, and compare this to the spectrum in our best fit annihilating dark matter (plus diffuse and HESS point source backgrounds) model, including the effect of the FGST's point spread function (which noticeably suppresses the lowest energy emission in the left frame). The distinctive bump-like feature observed at $\sim$1-5 GeV is easily accommodate by a fairly light dark matter particle ($m_{\rm dm} \approx 25-30$ GeV) which annihilates to a $b \bar{b}$ final state. The normalization was best fit to an annihilation cross section of $\sigma v \approx 9 \times 10^{-26}$ cm$^3$/s, which is a factor of about three times larger than that predicted for a simple $s$-wave annihilating thermal relic.

It is interesting to note that the annihilation rate and halo profile shape found to best accommodate the FGST data here is very similar to that required to produce the excess synchrotron emission known as the ``WMAP haze''~\cite{wmaphaze}. To be produced by a dark matter particle as light as that described in this scenario, however, the observed hardness of the haze spectrum requires a fairly strong magnetic field in the region of the Galactic Center. It had been previously recognized that if the WMAP haze is the product of dark matter annihilations, then the FGST would likely be capable of identifying the corresponding gamma ray signal~\cite{Hooper:2007gi}.

The low mass and relatively large annihilation cross section required in this scenario are also encouraging for the prospects of other gamma ray searches for dark matter annihilation products, such as those observing dwarf galaxies and efforts to detect nearby subhalos.

In conclusion, we have studied the angular distribution and energy spectrum of gamma rays measured by the Fermi Gamma Ray Space Telescope in the region surrounding the Galactic Center, and find that this data is well described by a scenario in which a 25-30 GeV dark matter particle, distributed with a halo profile slightly steeper than NFW ($\gamma =1.1$), is annihilating with a cross section within a factor of a few of the value predicted for a thermal relic. 

It should be noted, however, that if astrophysical backgrounds exist with a similar spectral shape and morphology to those predicted for annihilating dark matter (a spectrum peaking at $\sim 1-3$ GeV, distributed with approximate spherical symmetry around the Galactic Center proportional to $r^{-2.2}$), the analysis performed here would not differentiate the resulting background from dark matter annihilation products. Gamma rays from pion decay taking place with a roughly spherically symmetric distribution around the Galactic Center, for example, could be difficult to distinguish. Further information will thus be required to determine the origin of these photons.

{\it Acknowledgements:} We would like to thank Doug Finkbeiner for his help in processing the FGST data. We would also like to thank Greg Dobler and Neal Weiner for their helpful comments. DH is supported by the US Department of Energy, including grant DE-FG02-95ER40896, and by NASA grant NAG5-10842.%

\end{document}